\documentclass[12pt]{article}
\pdfoutput=1
\usepackage{amsmath,amsfonts,graphicx,color,bbm,tikz,bm,setspace}
\usepackage{comment}
\usepackage[nosort]{cite}
\usepackage{subfigure}
\usetikzlibrary{calc,positioning}
\usetikzlibrary{patterns,arrows,decorations.pathreplacing}
\usepackage{caption}
\usepackage{ulem}
\tikzset{>=stealth}
\usepackage{lipsum}
\usepackage{tabularx}
\usepackage{fullpage}
\usepackage{hyperref}
\usepackage{bbold}
\usepackage{amsthm}
\theoremstyle{definition}

\usepackage{listings}
\usepackage{tcolorbox}

\newcommand{\mathsym}[1]{{}}
\newcommand{\unicode}[1]{{}}

\makeatletter\@addtoreset{equation}{section}\makeatother

\newcommand{\be}{\begin{equation}}
\newcommand{\ee}{\end{equation}}
\def\beq{\begin{equation}}
\def\eeq{\end{equation}}
\newcommand{\bea}{\begin{eqnarray}}
\newcommand{\eea}{\end{eqnarray}}

\newcommand{\bra}[1]{{\left< {#1} \right|}}
\newcommand{\ket}[1]{{\left| {#1} \right>}}

\renewcommand{\title}[1]{\vbox{\center\LARGE{#1}}\vspace{3mm}}
\renewcommand{\author}[1]{\vbox{\center{#1}}\vspace{3mm}}

\newcommand{\email}[1]{\vbox{\center\tt#1}\vspace{3mm}}

\begin{document}

\begin{center}
{\large {\bf  Toffoli gates solve the tetrahedron equations }}

\author{ Akash Sinha,$^a$ Pramod Padmanabhan,$^a$ Vladimir Korepin$^{b}$}

{$^a${\it School of Basic Sciences,\\ Indian Institute of Technology, Bhubaneswar, 752050, India}}
\vskip0.1cm
{$^b${\it C. N. Yang Institute for Theoretical Physics, \\ Stony Brook University, New York 11794, USA}}

\email{akash26121999@gmail.com, pramod23phys@gmail.com, vladimir.korepin@stonybrook.edu}

\vskip 0.5cm 

\end{center}


\abstract{
\noindent 
The circuit model of quantum computation can be interpreted as a scattering process. In particular, factorised scattering operators result in integrable quantum circuits that provide universal quantum computation and are potentially less noisy. These are realized through Yang-Baxter or 2-simplex operators. A natural question is to extend this construction to higher qubit gates, like the Toffoli gates, which also lead to universal quantum computation but with shallower circuits. We show that unitary families of such operators are constructed by the 3-dimensional generalizations of the Yang-Baxter operators known as tetrahedron or 3-simplex operators. The latter satisfy a spectral parameter-dependent tetrahedron equation. This construction goes through for $n$-Toffoli gates realized using $n$-simplex operators.

}


\section{Introduction}
\label{sec:Introduction}
Quantum computation is a physical process \cite{feynman2018feynman}. The quantum circuit can be interpreted as the evolution of an arbitrary number of qubits, with quantum gates governing each time step. This can also be viewed as a scattering process of the qubits with the corresponding quantum circuit as a scattering matrix \cite{gu1989one, bose2011quantum}. For particular systems the scattering matrix factorizes into 2-qubit gates, reminiscent of the factorization of the scattering matrix in integrable models \cite{Zamolodchikov1979FactorizedSI}. The resulting 2-qubit gates are in fact the $R$-matrices that solve the Yang-Baxter equation \cite{YangCN1967,BAXTER1972193}. These circuits are potentially less noisy due to the many conserved quantities preventing information loss. This idea has gained traction in the integrable model and quantum computation communities in the last two decades \cite{kauffman2004braiding, zhang2005yang, zhang2013integrable, lloyd2014universal,Zhang2021QuantumCU}.

It is well known that most 2-qubit gates achieve universal quantum computation \cite{deutsch1995universality, divincenzo1995two} which have also been experimentally realized \cite{monroe1995demonstration, plantenberg2007demonstration, isenhower2010demonstration,o2003demonstration}. The circuit depths using such gates can be reduced with higher qubit gates, like the 3-qubit Toffoli gate \cite{toffoli1981bicontinuous}. Additionally these gates also provide classical and quantum universality \cite{deutsch1989quantum, deutsch1995universality, chau1995simple, barenco1995elementary, shi2002both, aharonov2003simple}. For some potential experimental realizations of such gates, see \cite{fedorov2012implementation, li2024hardware}. The question arises if it is possible to obtain these gates through integrable systems. The corresponding models have to satisfy a higher dimensional version of the Yang-Baxter equation. The study of such equations was initiated in the three dimensional case, first formulated by Zamalodchikov. This is the tetrahedron equation \cite{Zamolodchikov1980TetrahedronOriginal,Zamolodchikov1981TetrahedronEA} which was subsequently generalized to higher dimensions \cite{MAILLET1989221,Bazhanov1982ConditionsOC}. 

A huge deterrent in using these equations, especially in the context quantum computing, has been the lack of appropriate solutions\footnote{However see \cite{padmanabhan2024solving} for a systematic construction of solutions using Clifford algebras.}. In this work we overcome this for the spectral parameter-dependent versions of the higher simplex equations. We show that by associating a spectral parameter, and subsequently an operator, to each site we can build simple solutions that solve the equations trivially. The operators are chosen from any algebraically closed set. Among the many options we choose this set as the group $SU(2)$ and show that the resulting tetrahedron and higher simplex operators contain unitary families of the Toffoli gate and their higher qubit generalizations. 

\section{Spectral parameter-dependent tetrahedron operators}
\label{sec:tetrahedron}
The tetrahedron equation of Zamalodchikov describes the scattering of straight strings or line segments extending the idea of the Yang-Baxter equation to three dimensions \cite{Zamolodchikov1980TetrahedronOriginal,Zamolodchikov1981TetrahedronEA}. 
It takes different forms depending on the labelling used to describe the scattering process \cite{Hietarinta_1994}. We will use the vertex form whose constant (spectral parameter independent) version is given by\footnote{Note that we use $T$ instead of the usual $R$ to denote the tetrahedron operator.} :
\begin{equation}\label{eq:3simplex}
    T_{123}T_{145}T_{246}T_{356} = T_{356}T_{246}T_{145}T_{123}.
\end{equation}
The indices label a local Hilbert space $V$ which we take to be $\mathbb{C}^2$ in this paper.
In the context of higher simplex equation this is also called the 3-simplex equation, with the operator $T$ denoting the 3-simplex or tetrahedron operator. Note that the six indices in this equation appear an equal number of times on both sides. This will help us generate several seemingly trivial solutions following the ideas used in the Yang-Baxter case \cite{padmanabhan2024integrability}. Before we write this down we specify the spectral-parameter dependent form of \eqref{eq:3simplex} :
\begin{eqnarray}\label{eq:3simplexSP}
    & & T_{123}\left(\mu_1, \mu_2, \mu_3\right)T_{145}\left(\mu_1, \mu_4, \mu_5\right)T_{246}\left(\mu_2, \mu_4, \mu_6\right)T_{356}\left(\mu_3, \mu_5, \mu_6\right) \nonumber \\
    & = & T_{356}\left(\mu_3, \mu_5, \mu_6\right)T_{246}\left(\mu_2, \mu_4, \mu_6\right)T_{145}\left(\mu_1, \mu_4, \mu_5\right)T_{123}\left(\mu_1, \mu_2, \mu_3\right),
\end{eqnarray}
where the $\mu_j\in\mathbb{C}$'s are the spectral parameters associated to the site $j$. To solve this equation we consider a set of operators parametrized by these spectral parameters :
\begin{equation}\label{eq:Qset}
    \mathcal{S} = \left\{Q(\mu)|\mu\in\mathbb{C}\right\}~;~\left[Q(\mu), Q(\nu) \right]\neq 0.
\end{equation}
We take this set to form a closed finite or infinite dimensional algebra. A possible choice for the $Q$'s are matrices.  
While an Abelian set produces trivial tetrahedron operators, more non-trivial solutions are obtained when the elements do not commute for the different parameters $\mu$.
Moreover the elements may or may not be invertible. With these assumptions we can write down a general solution of the spectral-parameter dependent tetrahedron equation \eqref{eq:3simplexSP} as :
\begin{eqnarray}\label{eq:genTetrahedronOp}
    T_{ijk}\left(\mu_{ijk}\right) & = & \mathbb{1} + \alpha_1~Q(\mu_i) + \alpha_2~Q(\mu_j) + \alpha_3~Q(\mu_k) \nonumber \\ 
    & + & \beta_1~Q(\mu_i)Q(\mu_j) + \beta_2~Q(\mu_j)Q(\mu_k) + \beta_3~Q(\mu_k)Q(\mu_i) \nonumber \\
    & + & \gamma~Q(\mu_i)Q(\mu_j)Q(\mu_k).
\end{eqnarray}
Note that we have introduced a shorthand notation compressing the tuple of spectral parameters: $$\mu_{ijk}\equiv\left(\mu_i, \mu_j, \mu_k\right).$$ The parameters $\alpha$'s, $\beta$'s and $\gamma$ are constant complex numbers. They are similar to coupling constants. This operator indeed satisfies \eqref{eq:3simplexSP} trivially. This is a consequence of associating a particular operator to every site $j$ and the fact that there are an equal number of each of the six indices on both sides of the tetrahedron equation. For arbitrary choices of the set $\mathcal{S}$, these operators may not be unitary even if they are invertible. Two remarks are in order at this point. 
\begin{enumerate}
    \item Yang-Baxter, tetrahedron and more generally higher simplex operators can broadly be classified into two types : {\it local} and {\it non-local} solutions. The latter occurs when the space, the operators are acting on, does not have a tensor product structure. An example of this is the Fibonacci braid group \cite{Kauffman2008TheFM,Jana2022TopologicalQC}. As far as we know there are no examples of such non-local solutions for the tetrahedron and higher simplex equations. The key difference between the two types of solutions can be illustrated with the braid group (Yang-Baxter operators) case. To determine a $N$-strand braid group $B_N$ we need to determine $N-1$ generators. In the case of local solutions it is enough to determine one of them. The rest are obtained by the use of the permutation operators that appropriately change the indices of this operator. For non-local solutions, there is no {\it a priori} relations between the different generators and so all of them have to be determined separately. These statements are easily adapted to the higher simplex cases as well.
    
    The form of the tetrahedron operator \eqref{eq:genTetrahedronOp} gives the impression that this is a non-local solution of the tetrahedron equation as clearly :
    \begin{equation}
        P_{24}P_{35}T_{123}\left(\mu_{123}\right)P_{35}P_{24}=T_{145}\left(\mu_{123}\right)\neq T_{145}\left(\mu_{145}\right).
    \end{equation}
    Here the $P_{ij}$'s are the permutation operators [swap gates]: $$P_{ij}=\frac{1}{2}\left({\mathbb 1}+X_iX_j+Y_iY_j+Z_iZ_j\right)~;~X,Y,Z~\textrm{are Pauli matrices}.$$ This is because the permutation operators only swap the indices or the space on which the operators $Q$ are acting and does not alter the spectral parameters associated to those sites. However we can construct {\it twisted} permutation operators that swap both site indices and the spectral parameters associated with them. They are defined by their action on local operators :
    \begin{equation}
        P_{ij}\left(\mu_i, \mu_j\right)Q_i(\mu_i)P_{ij}\left(\mu_i, \mu_j\right) = Q_j(\mu_j),
    \end{equation}
    and they satisfy the relations of the permutation group generators :
    \begin{eqnarray}
        P_{ij}\left(\mu_i, \mu_j\right)P_{jk}\left(\mu_j, \mu_k\right)P_{ij}\left(\mu_i, \mu_j\right) & = & P_{jk}\left(\mu_j, \mu_k\right)P_{ij}\left(\mu_i, \mu_j\right)P_{jk}\left(\mu_j, \mu_k\right), \nonumber \\
        P_{ij}\left(\mu_i, \mu_j\right)^2 & = & \mathbb{1}, \\
        P_{i,i+1}\left(\mu_i, \mu_{i+1}\right)P_{k,k+1}\left(\mu_k, \mu_{k+1}\right) & = &  P_{k,k+1}\left(\mu_k, \mu_{k+1}\right)P_{i,i+1}\left(\mu_i, \mu_{i+1}\right)~;~k>i+1. \nonumber
    \end{eqnarray}
    \item This operator also solves the edge form of the tetrahedron equation trivially :
    \begin{eqnarray}\label{eq:3simplexSPedge}
        & & T_{123}\left(\mu_{123}\right)T_{124}\left(\mu_{124}\right)T_{134}\left(\mu_{134}\right)T_{234}\left(\mu_{234}\right) \nonumber \\
        & = & T_{234}\left(\mu_{234}\right)T_{134}\left(\mu_{134}\right)T_{124}\left(\mu_{124}\right)T_{123}\left(\mu_{123}\right).
    \end{eqnarray}
\end{enumerate}

\section{Toffoli gates from tetrahedron operators}
\label{sec:Toffoli}
We can now write down the Toffoli gates as spectral parameter-dependent tetrahedron operators. To do this we choose the set $\mathcal{S}$ as the group $SU(2)$ and the operators $Q(\mu)$ as the group elements. With this the tetrahedron operator takes the form :  
\begin{eqnarray}\label{eq:SU2tetrahedron}
   & T_{ijk}\left(\mathbf{u}_{ijk}~;~\mathbf{\theta}_{ijk}\right)  =  \left(\frac{\mathbb{1}+\mathrm{i}R(\hat{u}_i;\theta_i)}{2}\right)\left(\frac{\mathbb{1}+\mathrm{i}R(\hat{u}_j;\theta_j)}{2}\right) +  \left(\frac{\mathbb{1}+R(\hat{u}_i;\theta_i)R(\hat{u}_j;\theta_j)}{2}\right)& \nonumber \\
    & +  e^{\mathrm{i}\alpha}\left(\frac{\mathbb{1}-\mathrm{i}R(\hat{u}_i;\theta_i)}{2}\right)\left(\frac{\mathbb{1}-\mathrm{i}R(\hat{u}_j;\theta_j)}{2}\right)\mathrm{i}R(\hat{u}_k;\theta_k)~;~ R(\hat{u};\theta) = e^{-\mathrm{i}\theta\mathbf{\sigma}.\hat{u}}, &
\end{eqnarray}
with $R$ being the $SU(2)$ operator associated to every site $j$. The operator tuple $\sigma\equiv\begin{pmatrix}
    X&Y&Z
\end{pmatrix}$ contain the three Pauli matrices. On each site there are three spectral parameters given by the two independent real components of the unit vector $\mathbf{\hat{u}}_j$ and the real angle $\theta_j$. Note that this operator can be obtained from \eqref{eq:genTetrahedronOp} for particular choices of the $\alpha$, $\beta$ and $\gamma$ parameters. For arbitrary rotation operators $R$ this operator is not unitary. This operator satisfies the spectral parameter-dependent tetrahedron equation :
\begin{eqnarray}
    & & T_{123}(\mathbf{u}_{123};\theta_{123})T_{145}(\mathbf{u}_{145};\theta_{145})T_{246}(\mathbf{u}_{246};\theta_{246})T_{356}(\mathbf{u}_{356};\theta_{356})\nonumber \\
    & = & T_{356}(\mathbf{u}_{356};\theta_{356})T_{246}(\mathbf{u}_{246};\theta_{246})T_{145}(\mathbf{u}_{145};\theta_{145})T_{123}(\mathbf{u}_{123};\theta_{123}).
\end{eqnarray}
By choosing the spectral parameters as
\begin{equation}
   \hat{u}_i= \hat{u}_j = \begin{pmatrix}
        0 & 0 & 1
        \end{pmatrix},~\hat{u}_k = \begin{pmatrix}
        1 & 0 & 0
    \end{pmatrix}~;~\theta_i=\theta_j=\theta_k = \frac{\pi}{2},
\end{equation}
we obtain a unitary family of Toffoli gates parametrized by $\alpha$ \cite{deutsch1989quantum}:
\begin{eqnarray}\label{eq:unitaryToffoli}
    T_\alpha & = & \left(\frac{\mathbb{1}+Z_1}{2}\right)\left(\frac{\mathbb{1}+Z_2}{2}\right) + \left(\frac{\mathbb{1}-Z_1Z_2}{2}\right) \nonumber \\
    & +& e^{\mathrm{i}\alpha} \left(\frac{\mathbb{1}-Z_1}{2}\right)\left(\frac{\mathbb{1}-Z_2}{2}\right)X_3.
\end{eqnarray}
This coincides with the usual Toffoli gate at $\alpha=0$. Note that this is also the $CCNOT$ gate.

We will now generalize the $SU(2)$ tetrahedron operator \eqref{eq:SU2tetrahedron} in a manner generalizing the notion of a Toffoli gate. As with the usual Toffoli gate there are two control qubits whose eigenvalues determine the action on the third qubit. The two control qubits are now eigenstates of two arbitrary $SU(2)$ elements with eigenvalues $e^{\pm{\rm i}\theta_j}$, where $j=1,2$ are the indices for the two qubits. When the two eigenvalues are $e^{{\rm i}\theta_j}$, the computational basis states on the third qubit, rotated by the corresponding $SU(2)$ element, get flipped. Such a gate is constructed by defining two projection operators as follows
\begin{eqnarray}
\Pi^{\pm}\left(\hat{u}_i;\theta_i\right)=\frac{R(\hat{u}_i;\theta_i)-e^{\pm{\rm i}\theta_i}}{e^{\mp{\rm i}\theta_i}-e^{\pm{\rm i}\theta_i}},~\Pi^{\pm}\left(\hat{u}_i;\theta_i\right)^{\dagger}=\Pi^{\pm}\left(\hat{u}_i;\theta_i\right).
\end{eqnarray}
Using these we construct the tetrahedron operator
\begin{eqnarray}\label{eq:genUnitaryToffoli}
T_{ijk}\left(\mathbf{u}_{ijk}~;~\mathbf{\theta}_{ijk}\right)={\rm 1}-\Pi_i^-\left(\hat{u}_i;\theta_i\right)\Pi_j^-\left(\hat{u}_j;\theta_j\right)\left({\rm 1}-\tilde{X}_k\left(\hat{u}_k;\theta_k\right)\right),
\end{eqnarray}
where we have introduced $\tilde{X}_k\left(\hat{u}_k;\theta_k\right):=R(\hat{u}_k;\theta_k)XR(\hat{u}_k;\theta_k)^{\dagger}$. It is easy to verify that, $T_{ijk}$ as defined above, is indeed unitary. This coincides with the usual Toffoli gate at 
\begin{equation}
   \hat{u}_i= \hat{u}_j = \begin{pmatrix}
        0 & 0 & 1
        \end{pmatrix};~\theta_i=\theta_j= \frac{\pi}{2},~\theta_k =0.
\end{equation}

The twisted permutation operators in this case are obtained from
\begin{eqnarray}
    P(\hat{u}_1, \theta_1;\hat{u}_2,\theta_2)=R_1(\hat{u}_1,\theta_1)R_2(\hat{u}_2,\theta_2) .P . R_1(\hat{u}_1,\theta_1)^{\dagger}R_2(\hat{u}_2,\theta_2)^{\dagger}
\end{eqnarray}
where $P$ is the usual permutation operator. Then it is easy to check that if we have 
\begin{eqnarray}
    {\cal M}_j(\hat{u}_j,\theta_j;{\cal D})=R_j(\hat{u}_j,\theta_j) .{\cal D}.R_j(\hat{u}_j,\theta_j)^{\dagger}~;~j=\{1,2\}
\end{eqnarray}
then the operator $P$ acts as the twisted permutation operator that exchanges both the site index and the associated spectral parameters
\begin{eqnarray}
   P(\hat{u}_1,\theta_1;\hat{u}_2,\theta_2).{\cal M}_1(\hat{u}_1,\theta_1;{\cal D}).P(\hat{u}_1,\theta_1;\hat{u}_2,\theta_2)={\cal M}_2(\hat{u}_2,\theta_2;{\cal D})
\end{eqnarray}
for arbitrary ${\cal D}$.

\section{Discussion}
\label{sec:discussion}
The Toffoli gate is universal for classical computation that is reversible \cite{fredkin1982conservative}. This is achieved by choosing the parameter $\alpha$ in \eqref{eq:unitaryToffoli} to be irrational multiples of $\pi$. This is a crucial ingredient for any universal quantum computer as we expect the latter to simulate all reversible classical computation. To realize universal quantum computation the Toffoli gates require assistance from single qubit gates like the Hadamard gate \cite{shi2002both,aharonov2003simple}. While this is possible with the spectral parameter-dependent tetrahedron operators constructed earlier, we note that this can also be obtained from constant tetrahedron solutions. To see this consider the operator
\begin{equation}\label{eq:constantTetrahedronSolution}
    \tilde{T} = \mathbb{1} - \frac{1}{4}\left(\mathbb{1}-Z\right)\otimes\left(\mathbb{1}-Z\right)\otimes\left(\mathbb{1}-Z\right).
\end{equation}
This solves the constant [independent of spectral parameters] tetrahedron equation \eqref{eq:3simplex}. The second term in the right hand side can be multiplied by an arbitrary complex number and the resulting operator continues to obey the constant tetrahedron equation. In fact any arbitrary complex linear combination of the identity operator and the projector $\ket{111}\bra{111}$ is a constant tetrahedron operator. However they are not unitary in general. Also projectors to the vector $\ket{1}$ can be replaced by  three different projectors to three different [unrelated] vectors in corresponding two-dimensional spaces. This reduces the tetrahedron operator to the one in \eqref{eq:genUnitaryToffoli} rotated by an appropriately transformed Hadamard. This operator will solve the spectral parameter-dependent tetrahedron equation of \eqref{eq:3simplexSP}.

The operator $ \tilde{T}$ \eqref{eq:constantTetrahedronSolution} is equivalent to the Toffoli gate by the local unitary
\begin{equation}
    U = \mathbb{1}\otimes\mathbb{1}\otimes H~;~H=\frac{1}{\sqrt{2}}\left( X+Z\right),
\end{equation}
where $H$ is the Hadamard gate acting on the third qubit\footnote{The Yang-Baxter operator $\mathbb{1} - \frac{1}{2}\left(\mathbb{1}-Z\right)\otimes\left(\mathbb{1}-Z\right)$, is rotated by $\mathbb{1}\otimes H$ to the 2-qubit $CNOT$ gate.}. To obtain the unitary family of Toffoli gates \eqref{eq:unitaryToffoli} we rotate the constant tetrahedron operator 
\begin{equation}\label{eq:constantTetrahedronSolutionalpha}
    \tilde{T}_\alpha = \mathbb{1} - \frac{1}{4}\left(\mathbb{1}-Z\right)\otimes\left(\mathbb{1}-Z\right)\otimes\left(\mathbb{1}-e^{{\rm i}\alpha}Z\right)
\end{equation}
with a Hadamard on the third qubit. We can further generalize this by introducing a diagonal unitary matrix ${\cal U}\left(\alpha,\beta\right)={\rm diagonal}\left[e^{{\rm i}\alpha},e^{{\rm i}\beta}\right]$ to modify the above expression as
\begin{eqnarray}
    \tilde{T}_{\alpha,\beta} = \mathbb{1} - \frac{1}{4}\left(\mathbb{1}-Z\right)\otimes\left(\mathbb{1}-Z\right)\otimes\left(\mathbb{1}-{\cal U}\left(\alpha,\beta\right)Z\right).
\end{eqnarray}
This trivially satisfies the constant tetrahedron equation and is unitary as can be verified easily. These equivalences show that it is possible to achieve universal quantum computation using a tetrahedron operator, with and without spectral parameters, and the single qubit Hadamard gate. 

The constructions to obtain both the constant and the spectral parameter-dependent tetrahedron operators can be extended to an arbitrary number of qubits. The so called $n$-Toffoli gates are also universal for both reversible classical computation and quantum computation. The latter with a bit of assistance from single qubit gates. All of these can be obtained from higher dimensional generalizations of the tetrahedron equations, called the $n$-simplex equations \cite{MAILLET1989221}. These generalizations will become obvious once we write down the answers to the 4-simplex case. The spectral parameter-dependent 4-simplex equation \cite{Bazhanov1982ConditionsOC} is given by 
\begin{eqnarray}\label{eq:4simplexSP}
    & T_{1234}\left(\mu_{1234}\right)T_{1567}\left(\mu_{1567}\right)T_{2589}\left(\mu_{2589}\right)T_{368,10}\left(\mu_{368,10}\right)T_{479,10}\left(\mu_{479,10}\right) & \nonumber \\
    & = T_{479,10}\left(\mu_{479,10}\right)T_{368,10}\left(\mu_{368,10}\right)T_{2589}\left(\mu_{2589}\right)T_{1567}\left(\mu_{1567}\right)T_{1234}\left(\mu_{1234}\right). &
\end{eqnarray}
To solve this we choose operators from the set $\mathcal{S}$ \eqref{eq:Qset} such that every site is associated with a spectral parameter. When the set is the group $SU(2)$, we obtain a 4-simplex operator analogous to the tetrahedron operator in \eqref{eq:SU2tetrahedron} :
\begin{eqnarray}\label{eq:SU2-4-simplex}
   & T_{ijkl}\left(\mathbf{u}_{ijkl}~;~\mathbf{\theta}_{ijkl}\right)  = \mathbb{1}-\left(\frac{\mathbb{1}-\mathrm{i}R(\hat{u}_i;\theta_i)}{2}\right)\left(\frac{\mathbb{1}-\mathrm{i}R(\hat{u}_j;\theta_j)}{2}\right)\left(\frac{\mathbb{1}-\mathrm{i}R(\hat{u}_k;\theta_k)}{2}\right) & \nonumber \\
    & +  e^{\mathrm{i}\alpha}\left(\frac{\mathbb{1}-\mathrm{i}R(\hat{u}_i;\theta_i)}{2}\right)\left(\frac{\mathbb{1}-\mathrm{i}R(\hat{u}_j;\theta_j)}{2}\right)\mathrm{i}R(\hat{u}_l;\theta_l)~;~ R(\hat{u};\theta) = e^{-\mathrm{i}\theta\mathbf{\sigma}.\hat{u}}. &
\end{eqnarray}
This reduces to the 4-Toffoli gate when
\begin{equation}
   \hat{u}_i= \hat{u}_j = \hat{u}_k=\begin{pmatrix}
        0 & 0 & 1
        \end{pmatrix},~\hat{u}_l = \begin{pmatrix}
        1 & 0 & 0
    \end{pmatrix}~;~\theta_i=\theta_j=\theta_k = \theta_l= \frac{\pi}{2}.
\end{equation}
the constant 4-simplex operator corresponding to \eqref{eq:constantTetrahedronSolution} can also be generalized to the $n$-qubit case in the obvious manner. 

To summarise we have shown that the Toffoli gate and its higher qubit generalizations can be obtained from tetrahedron operators and its higher dimensional generalizations. We expect the integrable models corresponding to these higher simplex equations \cite{kuniba2022quantum} to provide reliable platforms for universal quantum computation with lesser noise. The considerations here thus bring these two fields (quantum computation and higher dimensional integrability) together in a promising manner. For future work it would be interesting to study the efficiency of circuits built out of such higher simplex gates. Breaking down these higher simplex gates into smaller ones could also help us find new solutions to higher simplex equations.

\section*{Acknowledgments}
VK is funded by the U.S. Department of Energy, Office of Science, National Quantum Information Science Research Centers, Co-Design Center for Quantum Advantage under Contract No. DE-SC0012704.

\bibliographystyle{acm}
\normalem
\bibliography{refs}

\begin{thebibliography}{10}

\bibitem{aharonov2003simple}
{\sc Aharonov, D.}
\newblock A simple proof that toffoli and hadamard are quantum universal.
\newblock {\em arXiv preprint quant-ph/0301040\/} (2003).

\bibitem{barenco1995elementary}
{\sc Barenco, A., Bennett, C.~H., Cleve, R., DiVincenzo, D.~P., Margolus, N.,
  Shor, P., Sleator, T., Smolin, J.~A., and Weinfurter, H.}
\newblock Elementary gates for quantum computation.
\newblock {\em Physical review A 52}, 5 (1995), 3457.

\bibitem{BAXTER1972193}
{\sc Baxter, R.~J.}
\newblock Partition function of the eight-vertex lattice model.
\newblock {\em Annals of Physics 70}, 1 (1972), 193--228.

\bibitem{Bazhanov1982ConditionsOC}
{\sc Bazhanov, V.~V., and Stroganov, Y.~G.}
\newblock Conditions of commutativity of transfer matrices on a
  multidimensional lattice.
\newblock {\em Theoretical and Mathematical Physics 52\/} (1982), 685--691.

\bibitem{bose2011quantum}
{\sc Bose, S., and Korepin, V.}
\newblock Quantum gates between flying qubits via spin-independent scattering.
\newblock {\em arXiv preprint arXiv:1106.2329\/} (2011).

\bibitem{chau1995simple}
{\sc Chau, H., and Wilczek, F.}
\newblock Simple realization of the fredkin gate using a series of two-body
  operators.
\newblock {\em Physical review letters 75}, 4 (1995), 748.

\bibitem{deutsch1989quantum}
{\sc Deutsch, D.~E.}
\newblock Quantum computational networks.
\newblock {\em Proceedings of the royal society of London. A. mathematical and
  physical sciences 425}, 1868 (1989), 73--90.

\bibitem{deutsch1995universality}
{\sc Deutsch, D.~E., Barenco, A., and Ekert, A.}
\newblock Universality in quantum computation.
\newblock {\em Proceedings of the Royal Society of London. Series A:
  Mathematical and Physical Sciences 449}, 1937 (1995), 669--677.

\bibitem{divincenzo1995two}
{\sc DiVincenzo, D.~P.}
\newblock Two-bit gates are universal for quantum computation.
\newblock {\em Physical Review A 51}, 2 (1995), 1015.

\bibitem{fedorov2012implementation}
{\sc Fedorov, A., Steffen, L., Baur, M., da~Silva, M.~P., and Wallraff, A.}
\newblock Implementation of a toffoli gate with superconducting circuits.
\newblock {\em Nature 481}, 7380 (2012), 170--172.

\bibitem{feynman2018feynman}
{\sc Feynman, R.~P.}
\newblock {\em Feynman lectures on computation}.
\newblock CRC Press, 2018.

\bibitem{fredkin1982conservative}
{\sc Fredkin, E., and Toffoli, T.}
\newblock Conservative logic.
\newblock {\em International Journal of theoretical physics 21}, 3 (1982),
  219--253.

\bibitem{gu1989one}
{\sc Gu, C., and Yang, C.~N.}
\newblock A one-dimensional n fermion problem with factorized s matrix.

\bibitem{Hietarinta_1994}
{\sc Hietarinta, J.}
\newblock Labelling schemes for tetrahedron equations and dualities between
  them.
\newblock {\em Journal of Physics A: Mathematical and General 27}, 17 (Sept.
  1994), 5727–5748.

\bibitem{isenhower2010demonstration}
{\sc Isenhower, L., Urban, E., Zhang, X., Gill, A., Henage, T., Johnson, T.~A.,
  Walker, T., and Saffman, M.}
\newblock Demonstration of a neutral atom controlled-not quantum gate.
\newblock {\em Physical review letters 104}, 1 (2010), 010503.

\bibitem{Jana2022TopologicalQC}
{\sc Jana, I., Montorsi, F., Padmanabhan, P., and Trancanelli, D.}
\newblock Topological quantum computation on supersymmetric spin chains.
\newblock {\em Journal of High Energy Physics 2023\/} (2022), 1--45.

\bibitem{kauffman2004braiding}
{\sc Kauffman, L.~H., and Lomonaco, S.~J.}
\newblock Braiding operators are universal quantum gates.
\newblock {\em New Journal of Physics 6}, 1 (2004), 134.

\bibitem{Kauffman2008TheFM}
{\sc Kauffman, L.~H., and Lomonaco, S.~J.}
\newblock The fibonacci model and the temperley-lieb algebra.
\newblock In {\em Defense + Commercial Sensing\/} (2008).

\bibitem{kuniba2022quantum}
{\sc Kuniba, A.}
\newblock {\em Quantum groups in three-dimensional integrability}.
\newblock Springer Nature, 2022.

\bibitem{li2024hardware}
{\sc Li, X.-L., Tao, Z., Yi, K., Luo, K., Zhang, L., Zhou, Y., Liu, S., Yan,
  T., Chen, Y., and Yu, D.}
\newblock Hardware-efficient and fast three-qubit gate in superconducting
  quantum circuits.
\newblock {\em Frontiers of Physics 19}, 5 (2024), 1--7.

\bibitem{lloyd2014universal}
{\sc Lloyd, S., and Montangero, S.}
\newblock Universal quantum computation in integrable systems.
\newblock {\em arXiv:1407.6634\/} (2014).

\bibitem{MAILLET1989221}
{\sc Maillet, J.~M., and Nijhoff, F.}
\newblock The tetrahedron equation and the four-simplex equation.
\newblock {\em Physics Letters A 134}, 4 (1989), 221--228.

\bibitem{monroe1995demonstration}
{\sc Monroe, C., Meekhof, D.~M., King, B.~E., Itano, W.~M., and Wineland,
  D.~J.}
\newblock Demonstration of a fundamental quantum logic gate.
\newblock {\em Physical review letters 75}, 25 (1995), 4714.

\bibitem{o2003demonstration}
{\sc O'Brien, J.~L., Pryde, G.~J., White, A.~G., Ralph, T.~C., and Branning,
  D.}
\newblock Demonstration of an all-optical quantum controlled-not gate.
\newblock {\em Nature 426}, 6964 (2003), 264--267.

\bibitem{padmanabhan2024integrability}
{\sc Padmanabhan, P., Hao, K., and Korepin, V.}
\newblock Yang-baxter solutions from commuting operators.
\newblock {\em arXiv:2401.05662\/} (2024).

\bibitem{padmanabhan2024solving}
{\sc Padmanabhan, P., and Korepin, V.}
\newblock Solving the yang-baxter, tetrahedron and higher simplex equations
  using clifford algebras.
\newblock {\em arXiv:2404.11501\/} (2024).

\bibitem{plantenberg2007demonstration}
{\sc Plantenberg, J., De~Groot, P., Harmans, C., and Mooij, J.}
\newblock Demonstration of controlled-not quantum gates on a pair of
  superconducting quantum bits.
\newblock {\em Nature 447}, 7146 (2007), 836--839.

\bibitem{shi2002both}
{\sc Shi, Y.}
\newblock Both toffoli and controlled-not need little help to do universal
  quantum computation.
\newblock {\em arXiv preprint quant-ph/0205115\/} (2002).

\bibitem{toffoli1981bicontinuous}
{\sc Toffoli, T.}
\newblock Bicontinuous extensions of invertible combinatorial functions.
\newblock {\em Mathematical Systems Theory 14}, 1 (1981), 13--23.

\bibitem{YangCN1967}
{\sc Yang, C.~N.}
\newblock Some exact results for the many-body problem in one dimension with
  repulsive delta-function interaction.
\newblock {\em Phys. Rev. Lett. 19\/} (Dec 1967), 1312--1315.

\bibitem{Zamolodchikov1980TetrahedronOriginal}
{\sc Zamolodchikov, A.~B.}
\newblock Tetrahedra equations and integrable systems in three-dimensional
  space.
\newblock {\em Zh.Eksper.Teoret Fizika 79\/} (1980).

\bibitem{Zamolodchikov1981TetrahedronEA}
{\sc Zamolodchikov, A.~B.}
\newblock Tetrahedron equations and the relativistic s-matrix of
  straight-strings in 2+1-dimensions.
\newblock {\em Communications in Mathematical Physics 79\/} (1981), 489--505.

\bibitem{Zamolodchikov1979FactorizedSI}
{\sc Zamolodchikov, A.~B., and Zamolodchikov, A.~B.}
\newblock Factorized s-matrices in two dimensions as the exact solutions of
  certain relativistic quantum field theory models.
\newblock {\em Annals of Physics 120\/} (1979), 253--291.

\bibitem{zhang2013integrable}
{\sc Zhang, Y.}
\newblock Integrable quantum computation.
\newblock {\em Quantum information processing 12\/} (2013), 631--639.

\bibitem{zhang2005yang}
{\sc Zhang, Y., Kauffman, L.~H., and Ge, M.-L.}
\newblock Yang--baxterizations, universal quantum gates and hamiltonians.
\newblock {\em Quantum Information Processing 4\/} (2005), 159--197.

\bibitem{Zhang2021QuantumCU}
{\sc Zhang, Y., and ping Wu, K.}
\newblock Quantum computation using action variables.
\newblock {\em Quantum Information Processing 21\/} (2021).

\end{thebibliography}

\end{document}